\documentclass{elsart}

\usepackage{natbib}

 \usepackage{graphicx}
 \usepackage{epsfig}

\usepackage{amssymb}

\begin{document}

\begin{frontmatter}


 \title{The ALTCRISS project on board the International Space Station}

\author[0] {M. Casolino\corauthref{cor}},
\corauth[cor]{Corresponding author.} \ead{casolino@roma2.infn.it}
\author[0]{F. Altamura},
\author[0]{M. Minori},
\author[0]{P. Picozza},
\author[1]{C. Fuglesang},
\author[2]{A. Galper},
\author[2]{A. Popov},
\author[3]{V. Benghin},
\author[3]{V. M. Petrov},
\author[4]{A. Nagamatsu},
\author[5]{T. Berger},
\author[5]{G. Reitz},
\author[6]{M. Durante},
\author[6]{M. Pugliese},
\author[6]{V. Roca},
\author[7]{L. Sihver}
\author[8]{F. Cucinotta},
\author[8]{E. Semones},
\author[9]{M. Shavers},
\author[10]{V. Guarnieri},
\author[10]{C. Lobascio}
\author[11]{D. Castagnolo}
\author[11]{R. Fortezza}
\address[0]{INFN and University of Rome Tor Vergata, Department of Physics, Via della Ricerca Scientifica 1, 00133 Rome, Italy }
\address[1]{European Astronaut Centre, ESA, Cologne}
\address[2]{Moscow State Engineering Physics Institute, Moscow, Russia}
\address[3]{Institute for Biomedical Problems, Moscow, Russia}
\address[4]{Japan Aerospace Exploration Agency, Japan}
\address[5]{DLR, Aerospace Medicine, Radiation Biology, Cologne, Germany}
\address[6]{University Federico II, and INFN Napoli, Italy}
\address[7]{Department of Nuclear Chemistry, Chalmers University of Technology,  Gothenburg, Sweden}
\address[8]{National Aeronautics and Space Administration, Lyndon B Johnson Space Centre, Houston, TX,
USA}
\address[9]{Radiation Biophysics Laboratory, Wyle Laboratories, Houston, TX, USA}
\address[10]{Alcatel Alenia Space Italia, Torino, Italy}
\address[11]{Mars s.r.l. Naples, Italy}


\title{}


\author{}

\address{}

\begin{abstract}
The Altcriss project aims to perform a long term survey of the
  radiation environment on board the International
Space Station. Measurements are being performed with active and
passive devices in different locations and orientations of the
Russian segment of the station. The goal is to perform a detailed
evaluation  of  the differences in particle fluence and nuclear
composition  due to different shielding material and attitude of
the station. The Sileye-3/Alteino detector is used  to identify
nuclei up to Iron in the energy range above $\simeq $ 60 MeV/n.
Several passive dosimeters (TLDs, CR39) are also placed  in the
same location of Sileye-3  detector. Polyethylene shielding is
periodically interposed  in front of the detectors to evaluate the
effectiveness of shielding on the nuclear component of  the
cosmic radiation.  The project was submitted to ESA in reply to
the AO in the Life and Physical Science of 2004  and data taking
began in December 2005. Dosimeters and data cards are rotated every six
months: up to now three launches of dosimeters and data cards have
been performed  and have been returned with the end of expedition
12 and 13.
\end{abstract}

{\small Accepted for publication on Advances in Space Research,
http://dx.doi.org/10.1016/j.asr.2007.04.037}

\end{frontmatter}

\section{Introduction}
The International Space Station is a unique testbed for long term
human permanence in space and a proving ground for future missions
to the Moon and Mars.   With these missions it is expected that
human presence in space will increase
 both in the number of space travellers and   the duration  of  the missions\cite{1,2,5}.
 Therefore a detailed knowledge of the radiation field in space, its
 effects   on human physiology and the associated risk have been   and will be needed in mission planning.
  In addition to a detailed understanding of the biological effects on the human body\cite{3},
  this task requires precise measurements of the particles in the cosmic ray environment,
  their temporal variations due to solar modulation or Solar Particle Events and their orbital
  dependence (due to geomagnetic cutoff and trapped particles encountered in the South Atlantic Anomaly)
  as well as to how these particles interact with the hull of the station.
 Nuclear abundances and differential spectra of cosmic ray nuclei have been studied over a wide energy range
and in different points of the Heliosphere  by a number of
spacecrafts (e.g. ACE, Sampex, Ulysses). This has greatly advanced the
knowledge of the cosmic ray  composition and the physical
phenomena involved such as solar modulation, solar energetic
particles, trapping of particles in the radiation belts and so on.
   Cosmic rays, consisting of $99\%$ H and He ions and $1\%$
   heavier nuclei (which - however - represent  the dominant
component of the equivalent  dose\cite{nasa1}) interact with the
spacecraft material, producing
   secondary particles that result in modified nuclear abundances and energy spectrum\cite{newcuci}.
   For these reasons   the radiation environment on board the ISS is being
monitored by  a number of different detectors employing different
techniques\cite{reitz,saka,badmir,badcuc,yasud,reiact,badhw,reiss}.
   Due to the large inhomogeneity in the hull of the station, it is still an open question  to  assess
     {\it a priori} the cosmic ray flux and the corresponding dose rate in the different points of
   the station\cite{dea,fluk,sivert,desire}. Furthermore, to these uncertainties one must add  the effects of a densely ionizing
    field on the human body, such as  the  accurate estimation of the damage to   cells and
the associated risk induced by heavily charged radiation on
astronauts\cite{Durante,Durante1,Durante2}.
    Radiation evaluation and  protection in space is therefore an
interdisciplinary field, involving scientists from many areas,
such as cosmic ray physics, radiobiology, dosimetry and computer
science. \section{The Altcriss project}
  The Altcriss (Alteino
Long Term Monitoring of Cosmic Rays on the International Space
Station) project aims to perform a long term survey of the
radiation   and cosmic ray environment on board the ISS.  It was
submitted to ESA in response to the AO in Life and Physical
Science of 2004  with observations beginning at the end of  2005
(increment 12) and expected to continue for three years. This
experiment follows previous ones on Mir where relative nuclear
abundances and Light Flash perception\cite{mir,lf,Casolino}
measurements have been performed with similar silicon detector
based devices (Sileye-1 and Sileye-2). Previous measurements on ISS
with Sileye-3/Alteino  have been performed in 2002 and 2005 in the
framework of the first and second Italian-Soyuz Missions. In those
missions measurements were limited to the taxi flight duration
($<$ 10 days each) and to the Pirs module.  The main goals  of this
  project are:
\begin{itemize}
\item {\it Monitoring of long and short term solar modulation  of  cosmic rays.} The active nature of the
device allows to identify particles of galactic, trapped and solar
origin according to their position and temporal profile.
Observations are currently  being carried at solar minimum, going
toward solar maximum.
\item {\it Observations of Solar Particle Events.}  We expect
in three years\cite{sheax} about 10 events with an energy and fluence high enough
to reach the interior of the station and trigger our detector.
For these events we plan to observe the temporal profile and the
nuclear abundances.
\item {\it Survey of different locations of the ISS modules.} By relocating and rotating the instrument it is possible to study the differences in flux and nature of cosmic rays due
to the different shielding of the station material (hull, racks,
instruments etc.). Flux  is also dependent on station attitude and
orientation: currently several locations in the Pirs (Russian
docking)   and in the Service Modules (Central area, crew cabins)
have been studied. In the future it is planned to  make
measurements in the Columbus module and in the US section of the
station.
\item {\it Study of the effectiveness of shielding materials.} Different materials are being considered to reduce
the dose to the astronauts: the current approach in weight
effective shielding in space is to use low Z materials for their
higher stopping power and fragmentation cross-section of the
projectile. In this way it is possible to reduce the LET (Linear
Energy Transfer) and the quality factor of the radiation, thus
reducing the equivalent dose to the astronauts. Although several
steps are being taken in this direction (such as putting water
reserves in the crew quarters) the best materials from this
standpoint  are often not practical. For instance, liquid hydrogen
would be  the best shielding material but cannot be used for the
dangers involved in handling such a material. In the Altcriss
project we are currently employing two set of tiles to study the
effect of shielding on the nuclear radiation field:
    \begin{enumerate}
        \item Polyethylene  tiles.
        These  are similar to what currently is used in the crew cabin of  the US section of the ISS.
        These tiles are located on top and bottom of the bidirectional acceptance window of the detector
           to evaluate the effect of this material (for a thickness of $\simeq 5g/cm^2$) on the radiation and the nuclear
           abundances. Passive dosimeters are interposed between
           the detector and the shielding tile to compare the dose
           measured with TLD and CR-39 with active data coming
           from Sileye-3/Alteino.
           \item Multimaterial tiles. These tiles are  divided into four sections, each composed  of a different material: Polyethylene, Kevlar,
           Nextel/Capton Composite and  one section left empty as a
           reference. These tiles were used in 2005 in the
           framework of the second Italian Soyuz Mission. In the
           Altcriss project they have until now been used with
           passive dosimeters interposed between the shielding
           tiles to measure the radiation dose.
           \end{enumerate}
           \item {\it Comparison with other detectors.} Given the complexity of the radiation field in space, in order to build a comprehensive picture
           of the cosmic ray environment on board the ISS it is necessary to correlate the measurements obtained with Sileye-3/Alteino with other   detectors
           on board the station. To this purpose the device was located in the starboard cabin  close to the Matroska-R spherical phantom.
            Furthermore a cross-comparison measurement campaign with the ESA Matroska\cite{berger} facility is planned to be carried
           forth during expedition 15: in this case Sileye-3 will be placed at the same locations as the human phantom (but not at the same time) to have the
           exact comparison of the cosmic ray flux. Comparison of the nuclear abundances measured with NASA IV-CPDS will also be
           performed. To study the propagation of cosmic
rays in the Earth's magnetosphere and from the exterior to the interior of
the station  the data
           coming from the Pamela experiment\cite{picozza}, a satellite  borne cosmic ray detector placed in a 350*650 km, $70^o$ inclination  will be used.

\end{itemize}

\section{Sileye-3/Alteino characteristics}

\begin{figure}
\begin{center}
\includegraphics[scale=0.60]{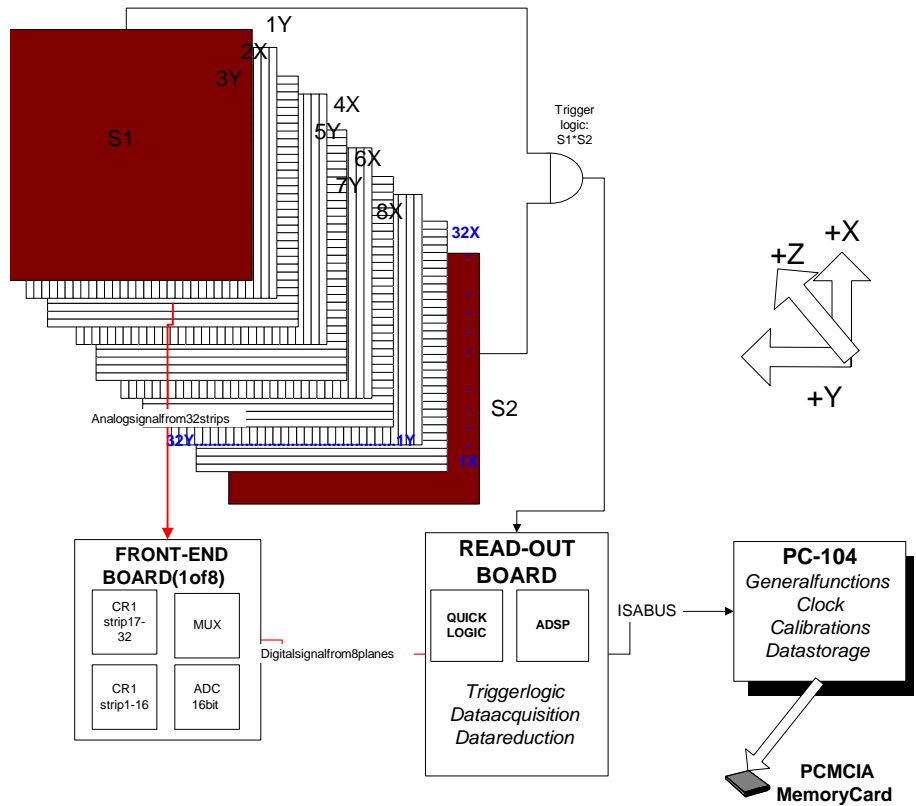}
\end{center}
 \caption{Scheme of Sileye-3 detector: the 8 silicon
planes are triggered by two scintillators (S1*S2), converted on
the front end boards (each with 2 CR1.4 preamplifier chip). The
signals are read by the DSP board which performs pedestal
subtraction and data compression. Data are then sent in blocks of
15kbyte  to the PC-104 CPU  via ISA bus to be saved on a PCMCIA
flash card.} \label{schema}
\end{figure}

Sileye-3 is a cosmic-ray detector composed of 8 silicon strip
detector planes, each divided in 32 strips, with 2.5 mm pitch.
There are 4 planes oriented along the X view and 4 planes along
the Y view. The general scheme of the detector is shown in Figure
\ref{schema}; a detailed description of the apparatus and its
functioning scheme can be found in \cite{elba,nara,cospar04}. The
device has a dynamic range capable of detecting particles from He
to above Iron. Also non-relativistic protons releasing a signal
above 1 mip\footnote{Minimum Ionizing Particle. It is defined as
the average energy lost in the detector by a minimum ionizing
($\simeq 2$ GeV proton)} can trigger the apparatus. Geometrical
factor is 23.78 $cm^2 sr$, considering that particles from both
sides can trigger the detector. Data are then stored in a
temporary buffer to be sent via an ISA interface to a storage and
data handling computer, a PC-104 board based on an AMD586 100 Mhz.
A quartz clock is used to save event time;  the beginning of each
session is written to synchronize (in the off-line stage) the
station time with the event time.   Data are stored on standard
PCMCIA cards. Their contents can  be downloaded to the ground via
telemetry, although usually only data samples relating to one day
of acquisition  are transferred using this procedure. The used
cards are sent  to the ground with the Soyuz and the ISS  crew at
the end of each increment. New cards  are uploaded with Soyuz or
Progress.

\section{Passive Dosimeters}

A number of passive dosimeters is used to measure the dose
absorbed in space in the shielded and unshielded configuration and
complement the active data coming from Sileye-3. These dosimeters
come from JAXA, DLR and Napoli Federico II University and consist
of different types of TLD and CR39 detectors. They are placed in
four pouches:
\begin{itemize}
\item  two pouches with all dosimeters  are interposed between the two
polyethylene shielding tiles and
 the acceptance windows of Sileye-3 when performing measurements in the shielded
 configuration (and thus are shielded by  roughly $2\pi$ polyethylene).
 When the Silicon detector is performing unshielded measurements the tiles and pouches are packed
 close one to the other and placed near the device (and the dosimeters are behind $4\pi $ polyethylene shielding).
 \item one pouch with four Federico II dosimeters is placed behind
 the multimaterial tile. Four samples of TLDs and CR39 are
 present in the unpackaged configuration   and are located  behind each material to facilitate the alignment
 between different materials and maximize the shielding geometrical
 factor.
 \item a control pouch with all dosimeters. This pouch is kept
 with data cards close to the detectors and moves to the different
 locations of the station.
 \item a ground control pouch  follows the others in all phases up  to launch in
 Baikonur.
 \end{itemize}

The pouches are rotated every 6 months, with each taxi flight;
except for the  first set of material that was launched at the end of
December and returned in April the duration was shorter.

\section{Survey of the radiation environment in the ISS}
\begin{figure}
\begin{center}
\includegraphics[scale=0.60]{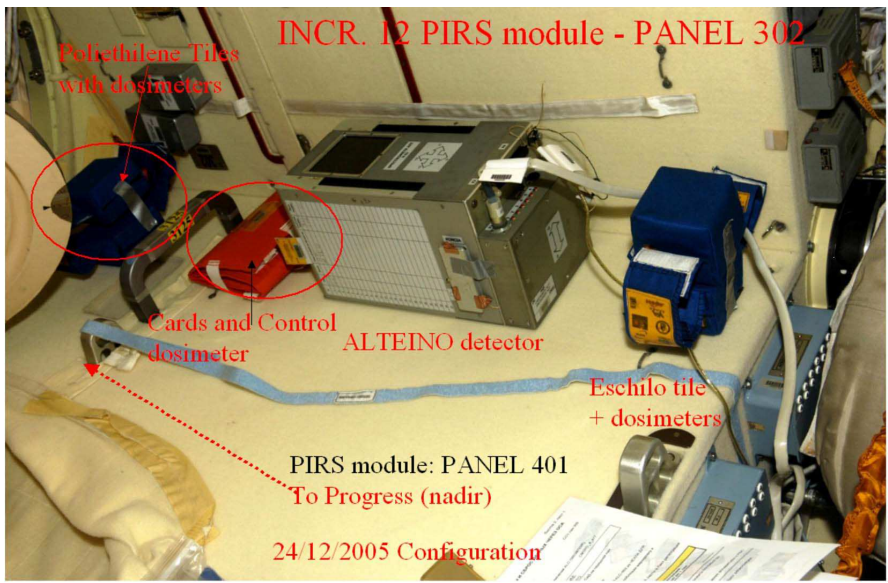}
 \end{center}
\caption{  Sileye-3   in the Pirs module (Panel 302) of the ISS in
its first operational position on of the Altcriss Survey
(24-12-2005). The device is in its unshielded configuration, with
the polyethylene shielding tiles on the left of the picture (in
the Soyuz / nadir direction) and the multimaterial tiles on the
right of the picture (in the direction of the station).}
\label{foto4}
\end{figure}

Data cards, dosimeters and polyethylene shielding necessary for
the experiment were first sent on board  ISS  on 21-12-05 with a
Progress craft. The detector was switched on 24-12-05 in the Pirs
module in the unshielded configuration. A first data sample of 40
hours was downlinked to the ground to verify the correct
functioning of the device. Subsequently two long term sessions
with and without shielding material (respectively 11 and 15 days)
in the Pirs module were performed. In January 2006 the measurement
campaign in the Russian Service module started: up to now the
device  has been located in both cabins and in several locations
of the  main area. In Figure \ref{foto4} is  shown the setup of
the experiment in the  Pirs/Docking module
 where  the detector was first located. It is possible to see the
 active detector and the shielding material pouches containing  the
 dosimeter bags.
For each position it has been tried (keeping into account all
constraints of logistics and observational time) to have a
shielded and an unshielded measurement; acquisitions with
different orientations at the same location have also been
performed to assess the differences in flux and nuclear abundances
due to different shielding material.

\section{Flight data and nuclear identification capabilities}
\begin{figure}
\begin{center}
\includegraphics[scale=0.60]{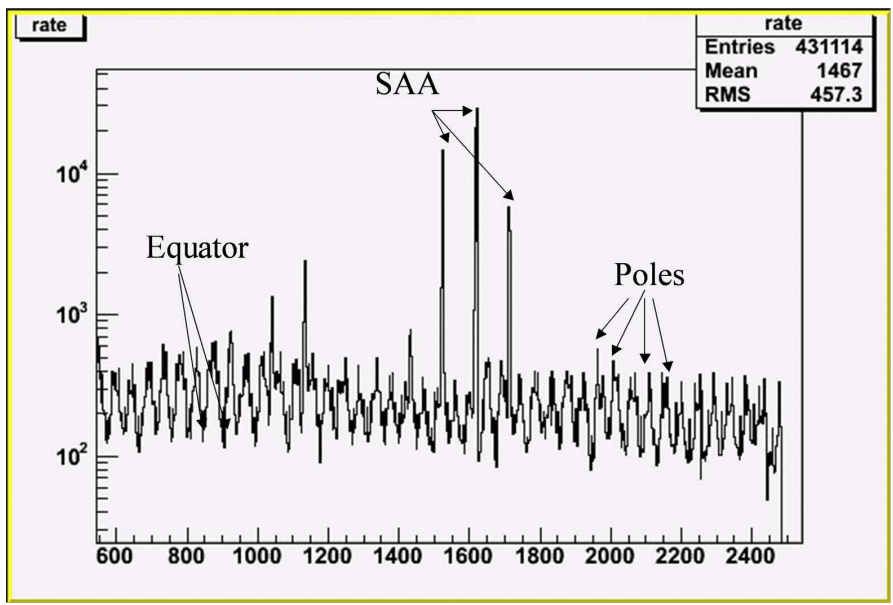}
\end{center}
 \caption{Cosmic ray flux (counts per minute) vs time (minutes) during  one typical acquisition session.
 Note the passage in the SAA (higher peaks) and the
modulation due to passage through the equator and the poles. }
\label{flussoshort}
\end{figure}

\begin{figure}
\begin{center}
\includegraphics[scale=0.60]{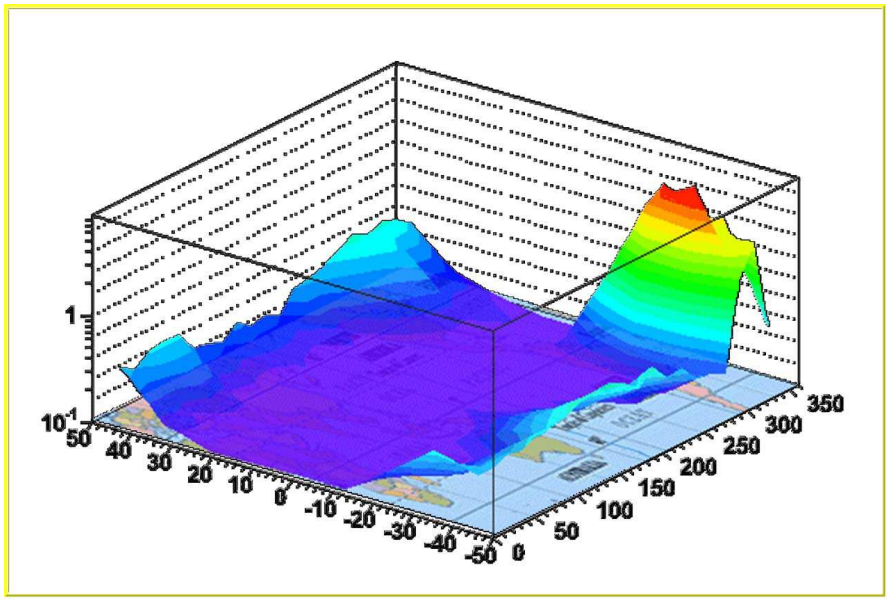}
 \end{center}
 \caption{All particle rate (arb. units) vs position measured with Sileye-3. It is possible to see the trapped proton peak in the South Atlantic Anomaly and the increase
 in the high latitude regions due to galactic nuclei. } \label{saa}
\end{figure}

In Figure \ref{flussoshort} is shown the   acquisition event  rate
as a function of time for a sample  of the dataset. Flux
modulation is due to the geomagnetic shielding, with higher rate
at the poles, where the cutoff is lower and lower rate at the
equator where the shielding is higher. The highest
  peaks occur during  passage in the South Atlantic Anomaly
(SAA), where particle rate increases more than one order of
magnitude due to the trapped proton component. Correlating
particle flux with position information it is possible to build an
all particle map (see Figure \ref{saa}) which shows the latitude
increase at high latitude due to galactic particles and the SAA
peak due to trapped protons.

\begin{figure}
\begin{center}
\includegraphics[scale=0.60]{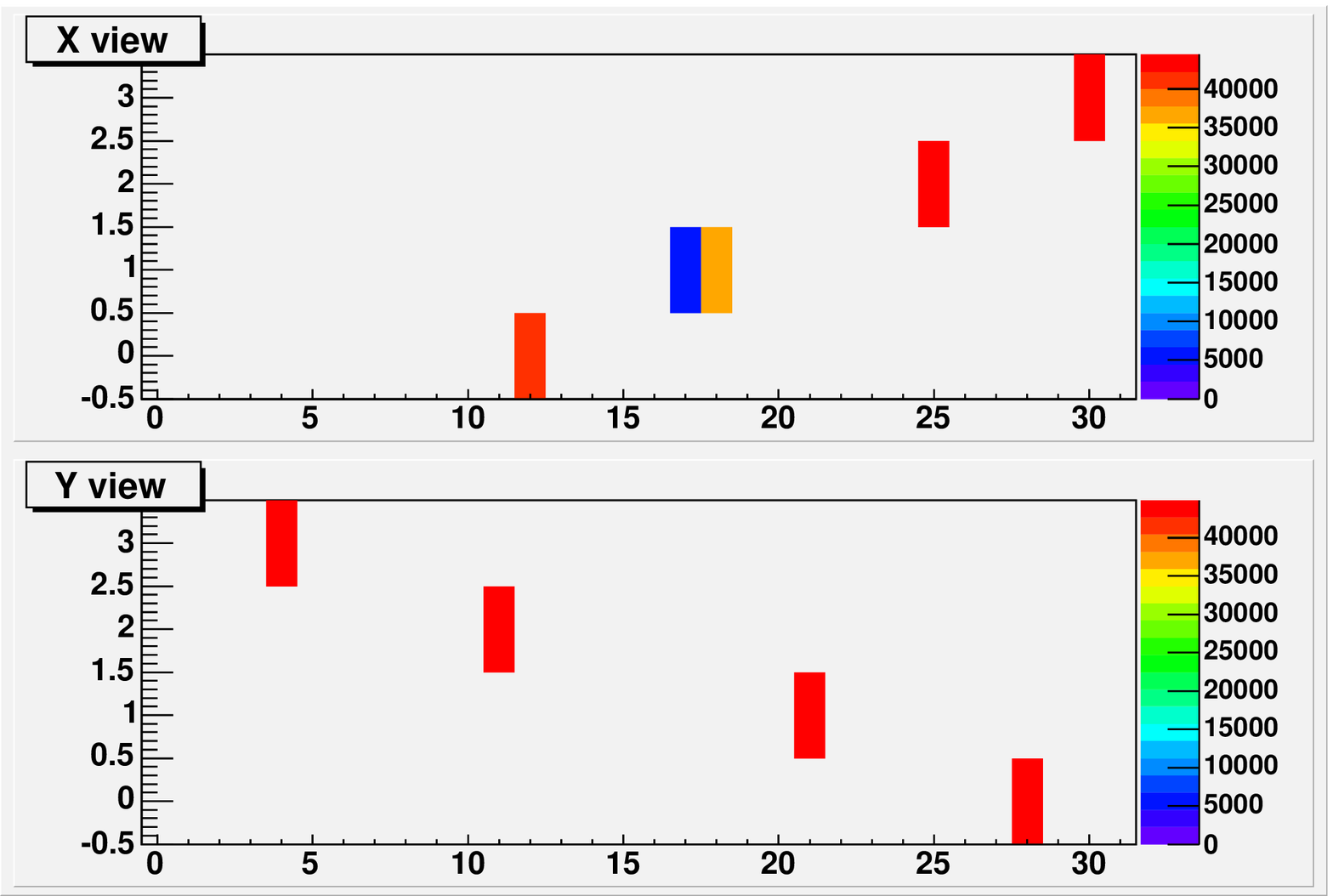}
\includegraphics[scale=0.45,angle=-90]{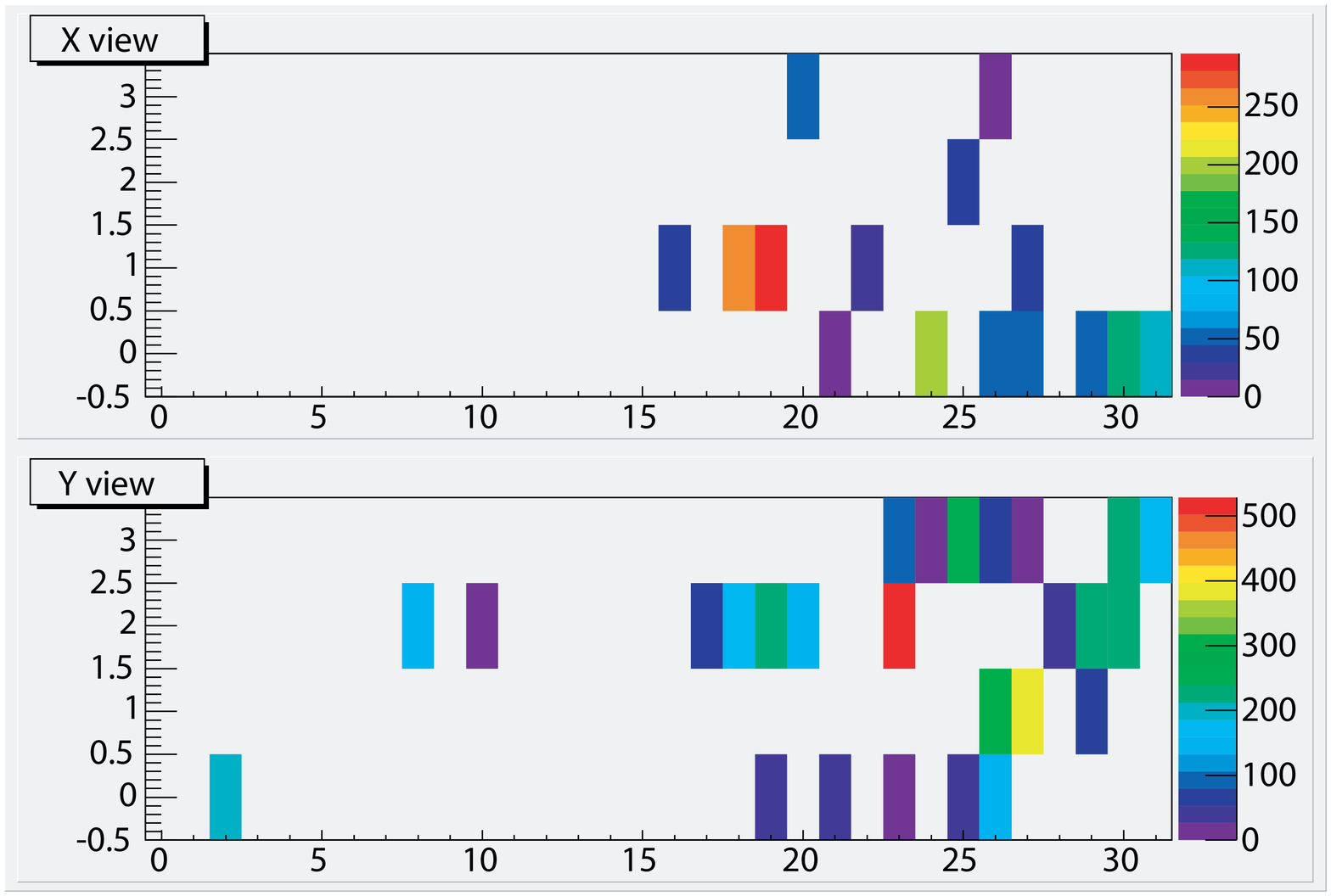}
\end{center}
\caption{Typical events obtained with Sileye-3. a: A  Neon
nucleus. b: A shower event crossing the detector. Top Panel: X
view, Bottom Panel Y view. x axis: strip number (1-32), y axis:
plane number (1-4). The energy released (in ADC channels) is shown
on the color bar on the right. Note the higher  energy deposited
in the Neon nucleus in respect to the shower
event\cite{cospar04}.} \label{evento}
\end{figure}

 A typical cosmic-ray event, in this case  a neon nucleus crossing the device,  is shown in
Figure \ref{evento} (Top). One of the characteristics of Sileye-3
is its independent channel readout which allows acquisition of
multi-particle events due to showers initiated with the
interaction of primary particles with the hull or the equipment of
the ISS: Figure \ref{evento} (Bottom) shows a typical event with
this topology.
To identify nuclei it is necessary to  select
single  tracks crossing the eight planes of the detector (noise is
removed from the events). The energy lost $E_{loss}$ in silicon
has been normalized to vertical incidence $E_{loss\: norm.}$
according to the formula: $E_{loss\: norm.}=E_{loss}
cos(\theta_{inc})$, with $\theta_{inc}$ the angle of incidence
from the normal of the silicon planes. An additional cut,
requiring that the energy released in the first and the last
planes does not
 differ by more than $20\%$,  selects particles of  energy $E_{kin}\gtrsim 70\: MeV/n$ (for Carbon, increasing with Z up to 150 MeV/n for Fe) .
Since the average energy loss of   nuclei  in matter is described
by the Bethe
    Block formula
    it basically depends from $Z^2/\beta^2$, with
    $\beta=v/c$ and $v$ the velocity of the impinging particle.
Given the differential energy spectrum of cosmic rays hitting the
detector, the value of $\beta $ (and thus the energy lost in the
detector) of most particles selected with the above mentioned cut
 is close to 1, with a smaller\footnote{For instance,
    $84\%$ of Carbon nuclei in cosmic rays have
    $1/\beta^2<1.2$.} amount of low energy particles  with a lower $\beta$,    resulting  in a higher energy release tail enlarging  the
 sigma of the gaussian peak. The peak  is further spread
    according to  the Landau distribution, which describes the
    distribution     of energy loss in matter.
Using the calibration obtained in \cite{cospar04}   particle
charge Z is obtained according to the following formula:
\begin{equation}
Z=\sqrt{\frac{E_{loss}(ADC) +3211.2}{371.04}}
\end{equation}

\begin{figure}
\begin{center}
\includegraphics[scale=0.60]{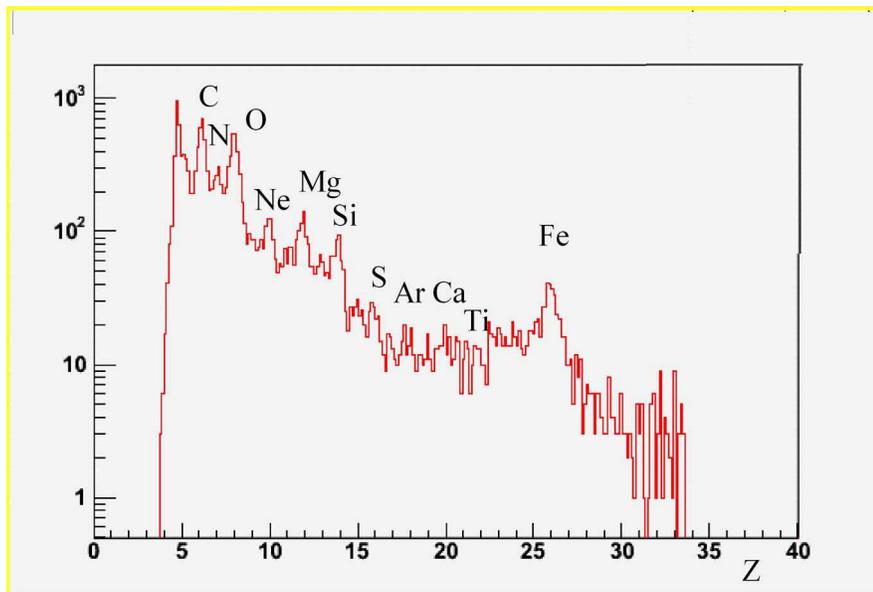}
\end{center}
 \caption{Histogram of particle counts showing the nuclear identification capabilities of Sileye-3 from C to Fe in the Pirs module. Note how  even numbered nuclei
 are more abundant than odd numbered ones. } \label{abbondanze}
\end{figure}

It is therefore possible to derive the   spectrum for nuclei up to
and above Iron, shown in Figure \ref{abbondanze}, in this case
referring to 11 days of data taking in the unshielded
configuration in the Pirs Module. It is possible to
    distinguish the peaks from C to Fe (sigma of gaussian fit for C is 0.3 charge units), with the even  Z  nuclei more abundant and evident  than the odd,
     as found in cosmic rays. Nuclear abundances and  trigger
     efficiency   in the different configurations is currently being
     evaluated. The active nature of the device allows for charge
     determination to be performed in different points of the
     orbit and different geomagnetic cutoff regions. Also
     abundance comparison spectra with/without  shielding will contribute to the
     determination of  effectiveness  of shielding materials in space.

\section{Conclusions}

In this work we have outlined the primary goals and presented
preliminary results of the Altcriss project on board the
International Space Station.    The data  gathered up to now are
under analysis and  will be useful for determining of the
radiation environment on board the station and the validation of
Montecarlo transport codes.   These measurements will be compared
with those obtained with   a large area detector, the Altea
facility\cite{altea},    sent to the ISS on July 2006. Altea will
also continue  to investigate the LF phenomenon and the
cosmic ray radiation in space.

\section{Acknowledgements}

We wish to thank all ESA and Energia staff, in particular M.
Heppener, E. Istasse, C. Mirra, A. Petrivelli, I. Nikolaiev, A.
Savchenko and H. Stenuit for their invaluable support and help in
planning and execution of the experiment.

\newpage

\begin{table}
\centering
\begin{tabular}{|l|l|}
\hline Electronics Characteristics &  \\
\hline

ADC type & 16 bit  \\
Pedestal position & $\simeq 5000\: ADC \: ch$  \\

Conversion Factor & 3.3 keV/ADC $\:\:$  8.7 eV/($\mu m$ ADC)  \\

1 MIP (Minimum Ionizing Protons) & 33 ADC ch $\:\:$ 108.6 keV
(0.286
keV/$\mu m$)  \\

Maximum Energy Detectable & 1600 MIP $\simeq $   174
MeV (460 keV $\mu m$)  \\
\hline
\end{tabular}
\label{caratt} \caption{Energy range of the detector
electronics\cite{cospar04}.}
\end{table}

\newpage

\end{document}